\def\aa{{A\&A}}
\def\aas{{A\&AS}}
\def\aj{{AJ}}

\def\apj{{ApJ}}
\def\apjs{{ApJS}}

\def\mnras{{MNRAS}}

\documentclass[cup5b]{caps}
\usepackage{graphicx}
\usepackage{amssymb}
\usepackage{ociwsymp4e}  

\HeadText{A. I. Karakas and J.C. Lattanzio}  

\newcommand{\Msun}{\ensuremath{{\rm M}_{\odot}}}

\newcommand{\iso}[2]{\hbox{${}^{#1}{\rm #2}$}}

\begin{document}

\pagenumbering{arabic}

\author[]{A. I. KARAKAS and J.C. LATTANZIO\\Centre for Stellar \& Planetary Astrophysics,
Monash University, Australia}

%
%

\chapter{Mg and Al Yields from Low and Intermediate Mass AGB stars}

\begin{abstract}

We investigate the production of aluminium and the heavy magnesium
isotopes in asymptotic giant branch (AGB) models. We evolve models with a
wide range in initial mass (1 $\geqslant (\Msun) \geqslant$ 6) and
composition ($Z=0.02, 0.008, 0.004$). We evolve the models from the
pre-main sequence, through all intermediate stages including the core
helium flash, to near the end of the thermally-pulsing AGB phase. We then
performed detailed nucleosynthesis calculations from which we determine
for the first time, the production of the magnesium and aluminium isotopes
as a function of the stellar mass and composition. From our models, we
calculate stellar yields suitable for galactic chemical evolution models.
We find that low-mass AGB stars ($M \lesssim 3\Msun$) do not produce the
necessary temperatures to synthesize the neutron-rich magnesium isotopes
in the helium shell.  The most massive AGB models do produce the
neutron-rich magnesium isotopes, and also produce \iso{26}Al in
substantial quantities via hot bottom burning. We note that the
calculations are subject to many uncertainties, such as the modelling of
the third dredge up, mass-loss and nuclear reaction rates.

\end{abstract}

\section{Introduction}

        In recent years our attempts to understand many aspects of
nucleosynthesis and stellar evolution have come to rely on our
understanding of the production of the magnesium and aluminium isotopes.
For example, abundance anomalies in globular cluster stars have been a problem for many years,
and the role of Mg and Al is central, and far from understood (Yong et al. 2003,
Shetrone 1996). At the heart of this problem is the quest for the production site
of the Mg and Al anomalies: are they produced in the star itself, and
mixed to the surface by some form of deep mixing (Denissenkov \& Weiss 1996)
or are they the result of pollution from an earlier generation of stars?
The latter would seem to implicate asymptotic giant branch (AGB) stars
(Denissenkov et al 1996, Ventura et al 2001) where Mg and Al can be produced
by thermal pulses and mixed into the envelope by the subsequent third dredge-up (TDU).

  	Although all isotopes of magnesium are produced by supernovae,
low-metallicity supernovae models fail to produce enough of the
neutron-rich Mg isotopes to account for the chemical evolution of
\iso{25}Mg and \iso{26}Mg in the Galaxy (Gay \& Lambert 2000).  Other
possible sources of the neutron-rich magnesium isotopes include the winds
from low-metallicity Wolf-Rayet (WR) (Maeder 1983)  and AGB stars (Siess
et al. 2002, Forestini \& Charbonnel 1997). There are currently no
quantitative studies of the production of the neutron-rich Mg isotopes in
low-metallicity WR stars. There are quantitative studies of magnesium
production in low metallicity AGB stars (Siess et al. 2002, Forestini \&
Charbonnel 1997)  but these studies do not cover a sufficiently large
range of mass or composition to produce yields suitable for galactic
chemical evolution models. For this reason, a quantitative understanding
of the production of the heavy Mg isotopes in AGB models of different mass
and metallicity is required if we are to understand the non-solar Mg
isotopic distribution observed in various stars.  For example, giants in
the globular cluster NGC 6752 were observed by Yong et al. (2003) to have
highly non-solar Mg isotopic ratios, with a slight excess of \iso{26}Mg
over \iso{25}Mg.  As the observed stars do not exhibit the luminosity
variations expected if the abundance anomalies were produced internally,
it was assumed that the giants were polluted by an earlier generation of
stars. The authors concluded that the earlier generation of stars were
likely to be a population of intermediate mass, very low metallicity 
($Z\sim 0$) AGB stars.

 	Aluminium is produced at the expense of Mg by the Mg-Al chain, a process
which can produce substantial \iso{26}Al (Arnould et al. 1999). Whilst most of the 
\iso{26}Al observed in the Galaxy today probably originated in young massive stars 
(Prantzos 1993)
contributions from other sources such as classical novae and low and intermediate 
mass AGB stars might be important (Meynet 1994). Models of classical novae by 
Jos\'{e} \& Hernanz (1998)  find substantial \iso{26}Al production.
The production and destruction of \iso{26}Al in AGB
stars has been discussed in detail by Mowlavi \& Meynet (1999).  They found that hot
bottom burning (HBB) in massive
AGB stars could be an important source of \iso{26}Al. Nollett, Busso \& Wasserburg (2003) 
studied parameterized extra-mixing processes in low-mass AGB
models. They found that, depending on the mixing parameters used, \iso{26}Al can be
produced in sufficient quantities to explain the amount inferred to have been present
in some circumstellar oxide grains at the time of their formation.
Whilst these various studies suffer from many uncertainties, they make the point that
there may be many sources contributing to the \iso{26}Al in the Galaxy.

A quantitative study of the production of the heavy magnesium isotopes and aluminium
in AGB models is the main aim of this contributed paper.

\section{Stellar Models}

Models were calculated with the Mount Stromlo Stellar Structure code (Wood \& Zarro 1981,
Frost \& Lattanzio 1996)
updated to include the OPAL opacities of Iglesias \& Rogers (1996).
Mass loss was included using the prescription of Vassiliadis \& Wood (1993) 
but without the modification for $M$ greater than 2.5$\Msun$. 
We calculated model sequences for three
different initial compositions: $Z = 0.02, 0.008$ and $0.004$ over a 
range in mass $1 \leqslant M_{0} (\Msun) \leqslant 6$ where $M_{0}$ is the
initial stellar mass. Initial abundances for the CNO elements were taken
from Grevesse, Noels \& Sauval (1992) for the $Z=0.02$ models, and from 
Russell \& Dopita (1992) for the Large Magellanic Cloud compositions ($Z=0.008$) and 
Small Magellanic Cloud compositions ($Z=0.004$).

We use the standard mixing-length theory
for convection, with a mixing-length parameter $\alpha = l / H_P = 1.75$.
We find the convective boundary at the base of the 
outer envelope by searching for a neutral border to the convective zone, 
in the manner described in Frost \& Lattanzio (1996) and Karakas, Lattanzio
\& Pols (2002). We note that while this method does increase the efficiency of 
the TDU for low-mass models, we do not find any dredge-up for the $Z=0.02$
models with $M \leqslant 2.0\Msun$. Reaction rates used in the evolution
code were taken mostly from Caughlan \& Fowler (1988), but with updates
included in the nucleosynthesis calculations (see below).

We performed detailed nucleosynthesis calculations separately
using a post-processing nucleosynthesis code which includes 
time-dependent diffusive mixing, 506 reactions and 74 species up to sulphur. 
We also include a small neutron capture network based on the 
iron-peak elements.  The bulk of the 506 reaction
rates are from the Reaclib Data Tables, based on the 1991 updated version 
of the compilation by Thielemann, Arnould \& Truran (1991). We include
recent reaction rates for  $\alpha$, proton and neutron capture
reactions when available, as detailed in Lugaro (1998).

\section{Production of Mg and Al in AGB stars}

  \begin{figure}
    \centering
    \includegraphics[width=11cm,angle=0]{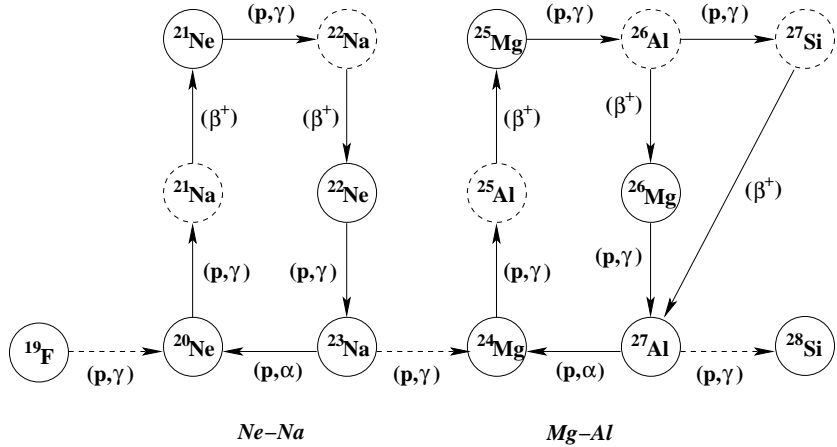}
    \caption{Reactions of the Ne-Na and Mg-Al chains. Unstable
    isotopes are denoted by dashed circles.}
    \label{fig:reactions}
  \end{figure}

The magnesium and aluminium isotopes are produced in three sites in AGB
stars: the hydrogen-burning shell (H-shell) via the Mg-Al chain, shown in
Fig.~\ref{fig:reactions} (Rolfs \& Rodney 1988); the helium-burning shell
(He-shell) via $\alpha$-capture on \iso{22}Ne and at the base of the
convective envelope in the most massive AGB stars that experience HBB,
again via the Mg-Al chain.  The efficiency of production of each site
depends in a complicated way on the temperature (i.e. via the initial
mass), initial composition and the extent to which each site affects the
other.

\subsection{Hydrogen-burning shell}

Magnesium and aluminium are produced in the H-burning shell via the 
activation of the Mg-Al chain.
In Fig.~\ref{fig:reactions} we show the reactions involved in the Ne-Na 
and the Mg-Al chains (Arnould et al. 1999, Rolfs \& Rodney 1988).
The first isotope in the Mg-Al chain to be affected is \iso{25}Mg,
which is burnt to \iso{26}Al when the temperature exceeds about
30 million K. The isotope \iso{26}Al is unstable to $\beta$-decay 
but the lifetime of $\beta$-decay relative to proton capture generally
favours proton capture within the H-burning shell. This produces 
the unstable \iso{27}Si which  $\beta$-decays (with a lifetime
on the order of a few seconds) to \iso{27}Al.
If temperatures  exceed 70~million~K, \iso{24}Mg  $+$ p leads to 
the production of \iso{25}Mg along with \iso{26}Al and \iso{27}Al.

In low-mass models\footnote{hereafter low-mass refers to models with $M \leqslant 2.5\Msun$},
the only change to the surface abundance of the Mg and Al isotopes comes 
from the H-burning shell. The ashes of the H-burning shell are first engulfed by 
the convective pocket before dredge-up occurs. In low-mass models, the Mg isotopes 
are not effected by He-shell burning but \iso{26}Al can be 
destroyed by neutron capture. Neutrons come from two reactions in
AGB stars: \iso{13}C($\alpha$, n)\iso{16}O and \iso{22}Ne($\alpha$,n)\iso{25}Mg.
As the temperature is too low for the activation of the \iso{22}Ne neutron source,
the only free neutrons are from the \iso{13}C neutron source. We do not include 
a \iso{13}C pocket in our models, so the neutrons in the convective pocket
are from the \iso{13}C left by the H-burning shell.

 \begin{figure}
   \centering
   \begin{tabular}{c}
    \includegraphics[width=8cm,angle=0]{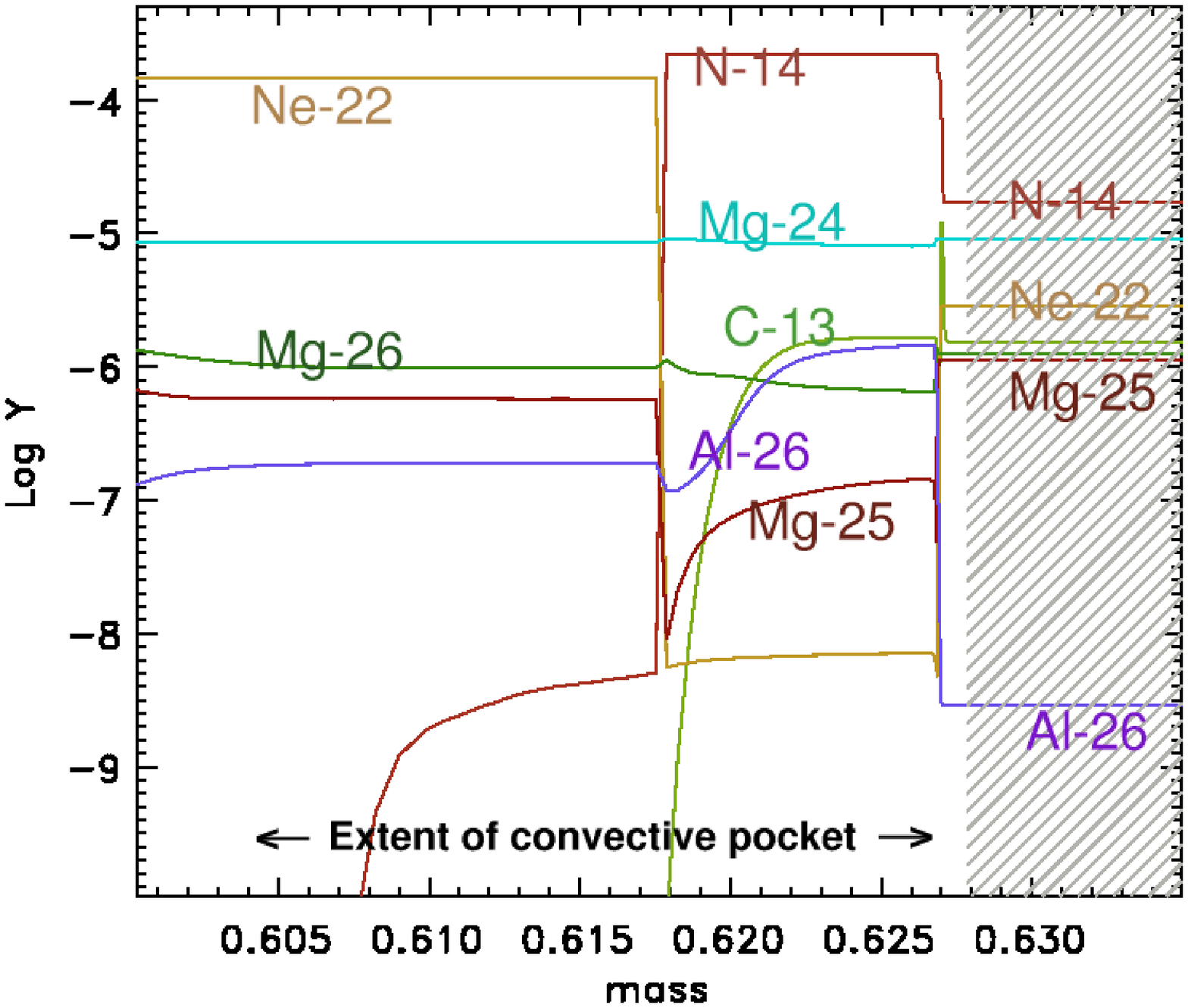} \\
    \includegraphics[width=8cm,angle=0]{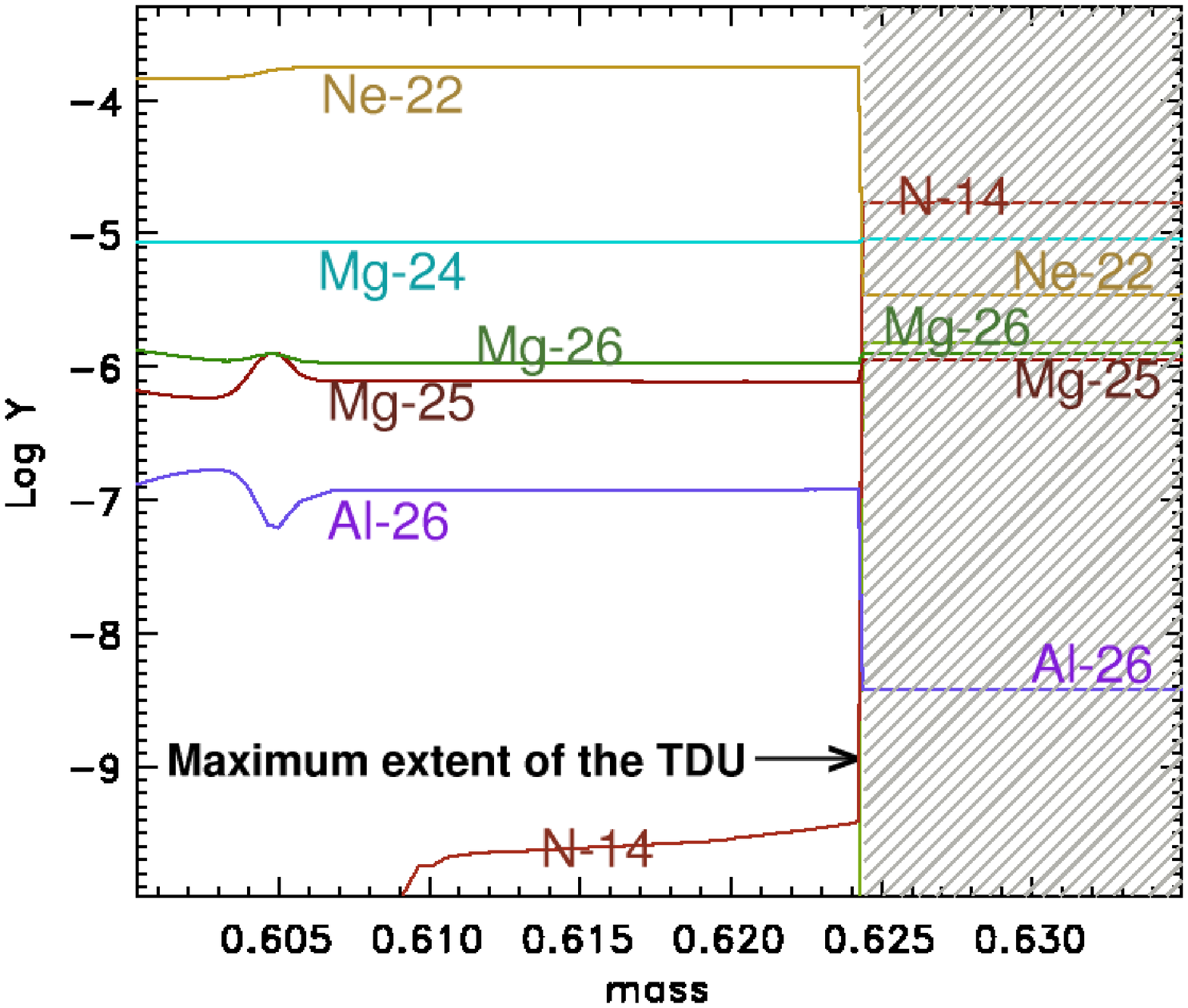} \\
   \end{tabular}
    \caption{Composition profile for the 1.5$\Msun$, $Z=0.004$ model
just before the 14$^{\rm th}$ thermal pulse (top panel) and at the maximum
extent of the following dredge up episode (lower panel).}
    \label{fig:m1.5z004}
  \end{figure}

We find that the change to the envelope composition in low-mass models
with efficient TDU is a slight depletion of \iso{25}Mg and a slight
increase in the abundance of \iso{26}Mg and \iso{27}Al.  The \iso{24}Mg
abundance remained unchanged. Owing to the lack of an efficient neutron
source in the low-mass models, we find that the surface abundance of
\iso{26}Al slowly increases with each dredge-up episode. By the end of the
TP-AGB phase, the \iso{26}Al/\iso{27}Al ratio could be as high as few
$\times 10^{-3}$ at the surface; except for the $Z=0.02$ models, where we
find this ratio to be about 100 times smaller. We demonstrate the effect
of H-burning nucleosynthesis in Fig.~\ref{fig:m1.5z004}. In the top panel
of Fig.~\ref{fig:m1.5z004} we plot the composition profile of the
1.5$\Msun$, $Z=0.004$ model just before the 14$^{\rm th}$ thermal pulse,
showing the He- and H-burning shells. The shaded region denotes the
convective envelope. The maximum extent of the convective pocket during
the 14$^{\rm th}$ thermal pulse is noted. In the lower panel of
Fig.~\ref{fig:m1.5z004} we plot the composition profile at the maximum
extent of the TDU, after the pulse. The composition of \iso{26}Al in the
intershell has been homogenized by the convective pocket, but is not
destroyed from neutron capture. We find that after dredge up occurs, the
surface abundance of \iso{26}Al has increased by about 30\%.

In conclusion, the operation of the H-shell in low-mass models is
quantitatively unimportant to the production of the Mg isotopes. Some \iso{26}Al
could be produced in low-mass, low-metallicity AGB models, but this conclusion 
suffers from many uncertainties.
As we discuss in the next two sections, the operation of the He-burning shell
and HBB is much more important in intermediate and massive AGB models than
the H-burning shell.

\subsection{Helium-burning shell}

The He-burning shell in AGB stars is a rich source of
nucleosynthesis. The main result is the production of \iso{12}C,
which when mixed to the surface may produce carbon stars.
There is also a wealth of other He-burning products such
as \iso{22}Ne, \iso{25}Mg, \iso{26}Mg (Forestini \& Charbonnel 1997) and s-process
elements (Busso et al. 2001).

  \begin{figure}
    \centering
    \includegraphics[width=8cm,angle=270]{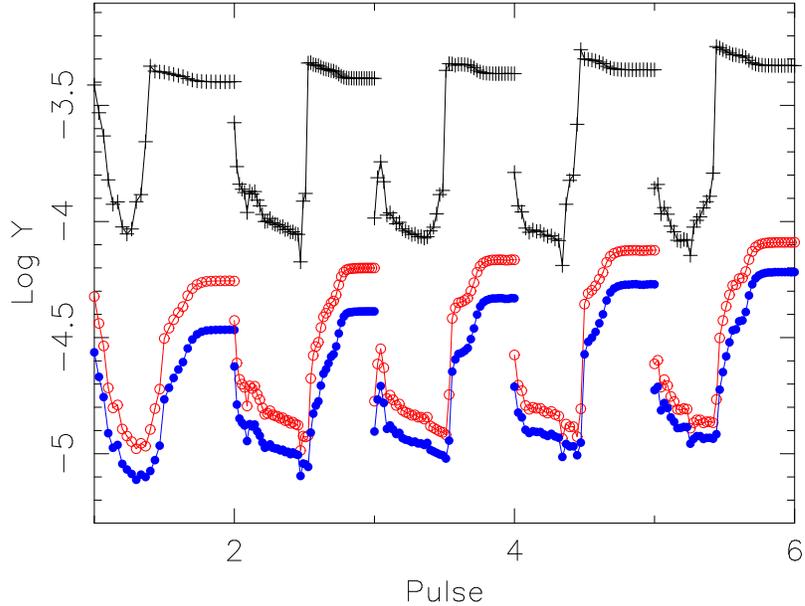}
    \caption{The intershell abundances of \iso{22}Ne (black plus signs), 
\iso{25}Mg (red open circles) and \iso{26}Mg (blue closed circles) 
as a function of pulse number. In this diagram we plot the intershell abundances for 
the 15th to the 20th pulse, but only during the time when the
convective shell is present; the $x$-axis is the (scaled) duration of the
convective pocket. Note that the abundances are the logarithm of the mole fraction, $Y$.}
    \label{fig:m4z008}
  \end{figure}

Substantial \iso{22}Ne is created during a thermal pulse by $\alpha$-capture 
onto the \iso{14}N left by the H-burning shell during the preceding 
interpulse period. If the temperature exceeds about 300~million~K, 
then \iso{25}Mg and \iso{26}Mg can be produced in  
substantial quantities by $\alpha$-capture onto 
\iso{22}Ne via the reactions \iso{22}Ne($\alpha$,n)\iso{25}Mg
and \iso{22}Ne($\alpha,\gamma$)\iso{26}Mg. In Fig.~\ref{fig:m4z008}
we plot the time variation of the intershell abundances of \iso{22}Ne, 
\iso{25}Mg and \iso{26}Mg for the 4$\Msun$, $Z=0.008$ model.
We plot the abundances in the intershell convective
region for the 15th to the 20th thermal pulse. The abundance for each species
initially decreases due to the growth of the convective shell
into the region previously processed by the H-shell. 
At the end of the preceding interpulse phase
this region has been depleted in \iso{22}Ne,  \iso{25}Mg and \iso{26}Mg via H burning.
As the temperature in the intershell convective 
region increases, successive $\alpha$-captures onto \iso{14}N first 
produces an increase in the \iso{22}Ne
abundance followed by an increase in \iso{25}Mg and \iso{26}Mg
when the temperature reaches exceeds 300~million~K. Note that after the
intershell convective pulse dies down, the
final \iso{22}Ne abundance is still high, making it the third 
most abundant species in this region 
(after He and \iso{12}C, but higher than \iso{16}O).

The exact amounts of \iso{25}Mg and \iso{26}Mg produced in the He-shell
is dependent not only on the reaction rates but also on the abundance
of matter left by the H-burning shell.  As the ashes of the H-shell are
engulfed by the next thermal pulse, the initial abundances of 
the two heavy magnesium isotopes can be quite different. For example,
in the 6$\Msun$, $Z=0.004$ model, we find that the abundance of
\iso{25}Mg/\iso{26}Mg can be as low as 0.2 at the beginning
of a thermal pulse (c.f. the initial ratio \iso{25}Mg/\iso{26}Mg $\sim$ 0.9). 
For this model, even if the temperature in the He-shell favours the production 
of \iso{25}Mg over \iso{26}Mg, we still find \iso{25}Mg/\iso{26}Mg $\sim$ 0.65
at the end of the thermal pulse (prior to TDU).

We find temperatures exceed 300~million~K in the He-shells of models with
$M \gtrsim 3\Msun$, depending on the initial composition. However we only find
substantial \iso{25}Mg and \iso{26}Mg production in the most massive
AGB models.
Thus we can conclude that the He-burning shell is the most efficient
production site of the neutron-rich Mg isotopes in AGB stars but only
in the most massive AGB models.

\subsection{Hot-bottom burning}

If the temperature at the base of the convective envelope
reaches about 60 million degrees K, {\em hot bottom burning} 
can occur, which is to say that the bottom of the convective envelope reaches 
into the top of the H-burning shell. We find H-burning primarily through the CNO 
cycle but also the Ne-Na and Mg-Al chains if the temperature is high enough.  
This site then becomes important for the production of many elements,
including primary nitrogen (Chieffi et al. 2001), lithium (Travaglio et
al 2001) and sodium (Mowlavi 1999).

  \begin{figure}
    \centering
    \includegraphics[width=10cm,angle=0]{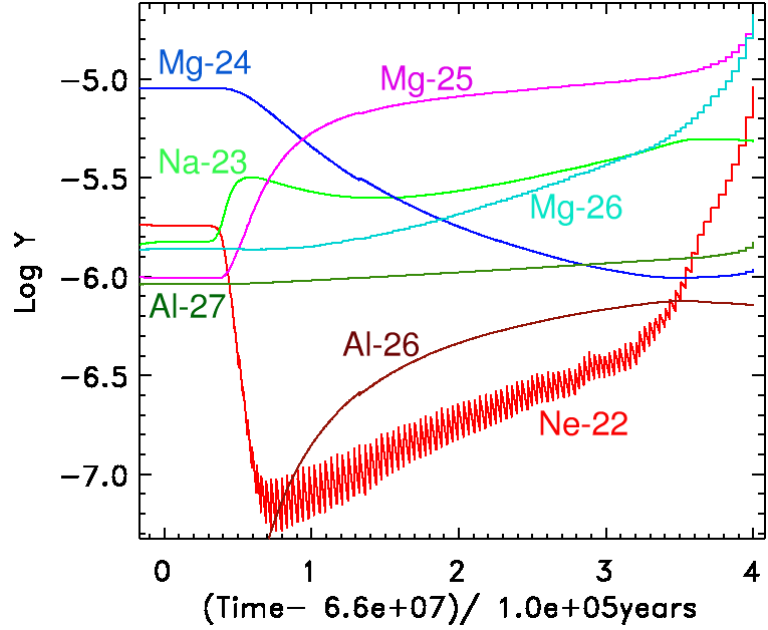}
    \caption{Surface abundance evolution during the AGB 
of the neon, sodium and magnesium isotopes 
for the the 6$\Msun$, $Z=0.004$ model.}
    \label{fig:m6z004-srf}
  \end{figure}

At the base of the convective envelope, the Mg-Al chain follows the 
same sequence as seen in the H-shell.  We note that although the region
hot enough for H-burning is quite thin, owing to efficient mixing the
entire envelope passes through the hot region at least 1000 times per interpulse.
In Fig.~\ref{fig:m6z004-srf} we plot the time variation of
some species at the surface of the 6$\Msun$, $Z=0.004$ model. This figure 
demonstrates the most extreme behaviour we found in the HBB models, 
with temperatures exceeding 94 million~K at the base of the 
convective envelope. We find large depletions in \iso{24}Mg and 
\iso{22}Ne followed by significant enhancements in \iso{25}Mg, 
\iso{26}Mg and \iso{26}Al. We also find moderate enhancements 
in \iso{23}Na and \iso{27}Al. This model was also significantly
depleted in \iso{16}O via HBB. After mass loss reduced the mass 
of the envelope below about 2$\Msun$, the temperature was too low 
for HBB and the continuation of 
dredge-up turned the model into an obscured carbon star, 
with C/O $\geqslant 1$ (see Frost et al 1998).

We conclude the HBB can be an efficient production site for \iso{26}Al
and \iso{27}Mg at the expense of the Mg isotopes. We note the most abundant
isotope, \iso{24}Mg is not burnt via HBB unless the temperature at the
base of the envelope exceeds about 80~million~K.

\section{Results and Discussion}

We calculate stellar yields according to the following definition:
\begin{equation}
 M_{\rm k} = \int_{0}^{\tau} \left[ X(k) - X_{0} (k)\right] \frac{d M}{dt} dt,
\label{eq:netyield}
\end{equation}
where $M_{ k}$ is the {\em net} yield of species $k$ (in solar masses),
$dM/dt$ is the current mass-loss rate, $X(k)$ and $X_{0} (k)$ refer
to the current and initial mass-fraction of species $k$ and $\tau$ is the
total lifetime of the stellar model.
The net yield can be negative, in the
case where the element is destroyed in the star and the final value
is lower than that during the main-sequence phase. A positive net
yield corresponds to those elements produced in the star so there
is a net enrichment over the stellar lifetime at the surface.

In practice, our  models does not lose the entire envelope during the
TP-AGB evolution owing to convergence difficulties near the end of the
AGB phase. For the lower masses considered, the remaining envelope mass 
is very small, and is certainly less than will be lost during the
subsequent interpulse phase. In these cases we calculate the yield
by simply removing the small remaining envelope with its current
composition. For the more massive models considered, there may be
enough envelope mass remaining for a few more thermal pulses to occur.
 HBB has been terminated,
however, so the species most affected are those which are present in the
intershell convective zone.
To calculate the stellar yields in these cases we will use
the principles of synthetic AGB evolution to calculate the enrichment
from the few remaining thermal pulses.
We do not go into details of the synthetic model but refer the
reader to Karakas \& Lattanzio (2003).

 \begin{figure}
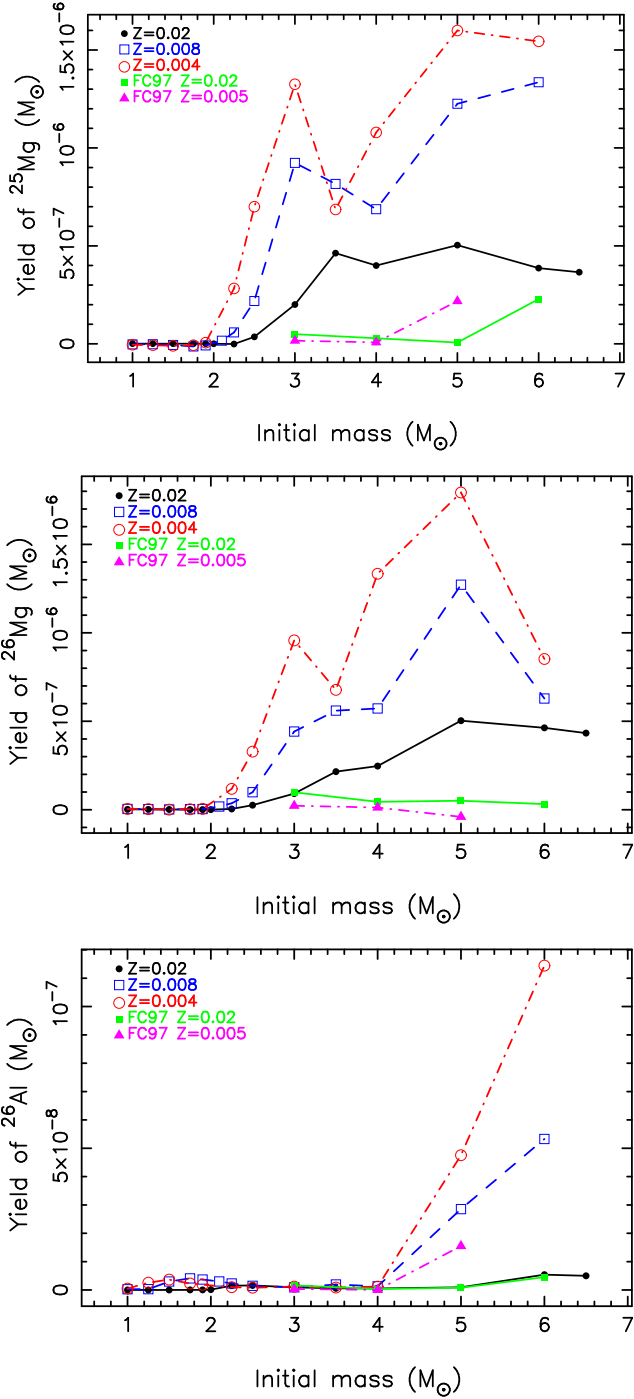

   \centering
   \begin{tabular}{c}
    \includegraphics[width=6cm,angle=270]{mg25-allz.ps} \\
    \includegraphics[width=6cm,angle=270]{mg26-allz.ps} \\
    \includegraphics[width=6cm,angle=270]{al26-allz.ps} 
   \end{tabular}
    \caption{Weighted yield of \iso{25}Mg (upper panel),
\iso{26}Mg (middle) and \iso{26}Al (lower panel) as a function
of the initial stellar mass (in $\Msun$). See text for a description
of the symbols.}
    \label{fig:yield}
  \end{figure}

In Fig.~\ref{fig:yield} we plot the yields of \iso{25}Mg (upper panel),
\iso{26}Mg (middle panel)  and \iso{26}Al (lower panel) as a function of
the stellar mass and composition. We weight the stellar yields by the
initial mass function of Kroupa, Tout \& Gilmore (1993). In each figure,
the black solid line (and points) refer to the $Z=0.02$ models, the blue
dashed line (and open squares) refer to the $Z=0.008$ models and the red
dot-dashed line (and open circles) refer to the $Z=0.004$ models. We plot
for comparison the yields of Forestini \& Charbonnel (1997) (hereafter
FC97), also weighted by the IMF.  The solid green squares are the $Z=0.02$
results from FC97 and the solid magenta triangles the $Z=0.005$ results
from FC97.  The first thing we note from Fig.~\ref{fig:yield} is that the
yields are highly metallicity dependent. For the three species considered,
the $Z=0.004$ yields are considerably larger than the $Z=0.02$ yields. The
yields are also highly dependent on the initial stellar mass. As expected,
low-mass models contribute little to the production of neutron-rich Mg
isotopes, \iso{25}Mg and \iso{26}Mg or to the radionuclide \iso{26}Al.  
Also as expected from the above discussion, the yields from the models
with HBB produce the largest amount of \iso{26}Al. If we compare our
results to FC97 we find we produce more of the neutron-rich Mg isotopes at
all masses and metallicities. We also produce more \iso{26}Al in the
$Z=0.008$ and $Z=0.004$ models but about the same at $Z=0.02$.

The large difference between our yields and those of FC97 is most likely
explained by the different modelling approaches used. We used detailed
stellar models for most of the TP-AGB phase, only resorting to synthetic
modelling for the final few thermal pulses. This means that we do not have
to treat HBB synthetically, as HBB had ended by the time the detailed
model calculations ceased. In comparison, FC97 use detailed modelling for
the pre-AGB phase and a few thermal pulses. The majority of the thermal
pulses, and the HBB phase, were calculated synthetically. The surface
abundance changes caused by HBB are highly dependent on the temperature
(and the density) at the base of the convective envelope. If these
quantities are not treated correctly in the synthetic model, the resulting
yields will be quite different to those found from detailed modelling. 
For example, FC97 extrapolated the behaviour of the temperature at the 
base of the envelope forward in time, realising that this extrapolation
was likely to be incorrect.
The differences between synthetic and detailed modelling also applies to 
those species affected by He-shell burning, such as the neutron-rich Mg
isotopes. A synthetic calculation can not follow the change to the
intershell composition with time. Indeed, most calculations (Marigo 2001)  
assume that the intershell composition remains constant over the entire
TP-AGB phase.  We find that the intershell abundance varies not only with
mass but also with time, and that the peak production of some species is
found right at the end of the TP-AGB, when the He-shell is hottest.

 We also note that the stellar yields are also dependent on the final
remnant mass. The final mass of a stellar model depends on the details of
the previous core H and He-burning phases as well as on the mass-loss
rates. These details can differ dramatically from one calculation to
another, making direct comparison difficult.

\section{Conclusions}

In conclusion, we find that intermediate mass TP-AGB models can produce substantial
quantities of the neutron-rich Mg isotopes from He-shell burning. The most massive
AGB models can also produce substantial \iso{26}Al from HBB. The yields presented here
are subject to many uncertainties, including the modelling of the third
dredge up as well as reaction rate uncertainties. 
Recent observations of non-solar Mg isotopic ratios could help constrain some
of these uncertainties. For example, Yong et al. (2003) found non-solar Mg isotopic
ratios in giant stars in the globular cluster stars NGC 6752, with a slight excess
of \iso{26}Mg over \iso{25}Mg in most of the stars. Yong et al. (2003)
discuss the possibility of AGB stars polluting the stars in this cluster. Whilst 
our more massive models ($\sim 5\Msun$) produce Mg isotopic ratios consistent with
their observations, most of our models have an excess of \iso{25}Mg over \iso{26}Mg.
Clearly further work needs to be done, including low-metallicity AGB models
and a detailed study of the dependence of mass-loss and the reaction rate 
uncertainties on the yields.

\end{document}